\documentclass[aps,pra,showpacs,preprint,amsmath,amssymb,groupedaddress]{revtex4-1}

\usepackage{graphicx}
\begin{document}

\title{Adiabatic approximation in the ultrastrong-coupling regime of a system consisting of an oscillator and two qubits}

\author{Ping Yang, Zhi-Ming Zhang}

\email[Corresponding author's email address: ]{zmzhang@scnu.edu.cn}

\affiliation{Laboratory of Nanophotonic Functional Materials and Devices, SIPSE and LQIT, South China Normal University, Guangzhou 510006, China}

\date{\today}

\begin{abstract}
We present a system composed of two flux qubits and a
transmission-line resonator. Instead of using the rotating wave
approximation (RWA), we analyse the system by the adiabatical
approximation methods under two opposite extreme conditions. Basic
properties of the system are calculated and compared under these two
different conditions. Energy-level spectrum of the system in the
adiabatical displaced oscillator basis is shown, and the theoretical
result is compared with the numerical solution.
\end{abstract}

 \pacs{42.50.-p, 42.50.Dv, 85.25.Cp}

\maketitle

\section{introduction}

Recently, more and more attention has been paid to superconducting
devices~\cite{Blais-2004-PRA,Feng-2012-PRA,Peropadre-2010-PRL,
Koch-2007-PRA,Haack-2010-PRB,Nataf-2010-PRL} to build quantum
systems, as the tunability of system parameters, which is one of the
most exciting advantages of superconducting circuit QED over
natural-particle-based cavity QED, makes such devices more likely to
be successful in quantum information
processing~\cite{Bushev-2011-PRB,Chen-2008-PRL,Yang-2004-PRL}. In
fact, some remarkable progresses have been made in recent years,
such as Fock states preparation in superconducting
devices~\cite{Hofheinz-2008-Nature}, single-photon router in the
microwave regime~\cite{Hoi-2011-PRL}, and high-fidelity readout in
circuit QED~\cite{Reed-2010-PRL}. Superconducting flux qubit that
consists of superconducting loops and Josephson junctions can be
viewed as a two-level system when the parameters satisfy the
condition called degeneracy point~\cite{Makhlin-2001-RMP}.

The paper is organized as follows: In Sec. II we introduce the
system and the Hamiltonian. In Sec. III we discuss the properties of
the system by adiabatic approximation under two opposite extreme
conditions. Energy-level spectrum of the system in the adiabatical
displaced oscillator basis is shown, and the theoretical result is
compared with the numerical solution. Sec. IV is the conclusion.

\section{system and hamiltonian}

The system that we consider consists of a harmonic oscillator and
two three-Josephson-junction qubits (The smallest junction has been
replaced by an additional loop ~\cite{Gustavsson-2011-PRA})
 which are coupled to the oscillator. Hamiltonian of such qubit is
given by $\hat{H}_{q_i}$ in Eq. (\ref{detail Hamiltonian}).

\begin{figure}[htb]
\centerline{\includegraphics[width=10cm]{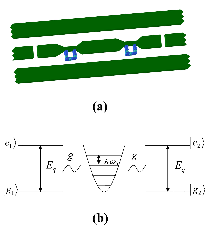}} \caption{(Color
online) (a) Schematic diagram of the structure. The two light blue
squares  are improved three-junction flux qubits fabricated to the
center conductor. (b) Schematic graph of the system. Two identical
qubits (i.e. parameters $\Delta$, $\epsilon$, energy-level splitting
$E_q$ and coupling strength $g$ for both qubits are of the same
value) viewed as two-level system with ground state $|g\rangle$ and
excited state $|e\rangle$, are coupled to a harmonic oscillator
whose characteristic frequency is $\omega_0 $. }\label{Fig1}
\end{figure}

The oscillator here is a microwave transmission-line resonator, and
the qubits which are superconducting flux qubits are fabricated so
that the loops are closed by the center
conductor~\cite{Bourassa-2009-PRA}. The schematic diagram of the
structure is shown in Fig. \ref{Fig1}(a), and it is worth noting
that the distance between the two qubits is sufficient large such
that the interaction between them can be ignored. The schematic
graph of our system is illustrated in Fig. \ref{Fig1}(b), in which
two qubits are coupled to a harmonic oscillator. It is
worth noting that the qubits are considered to be identical, which
means that the parameters $\Delta$, $\epsilon$, $E_q$, and the coupling
strength $g$ for both of the qubits are of the same value.

The Hamiltonian corresponding to our system is

\begin{equation}\label{total Hamiltonian}
\hat{H}=\sum_{i=1,2}\hat{H}_{q_i}+\hat{H}_{os}+\hat{H}_{int},
\end{equation}
where
\begin{equation}\label{detail Hamiltonian}
 \begin{aligned}
  \hat{H}_{q_i}=-\frac{\Delta}{2}\hat{\sigma}_{x_i}-\frac{\epsilon}{2}\hat{\sigma}_{z_i},  \\
  \hat{H}_{os}=\frac{\hat{p}^2}{2m}+\frac{1}{2}m\omega_0^2\hat{x}^2,  \\
  \hat{H}_{int}=g\hat{x}(\hat{\sigma}_{z_1}+\hat{\sigma}_{z_2}), \\
\end{aligned}
\end{equation}
where $\Delta$ is the energy gap tuned by the flux in the additional
loop, and $\epsilon$ is the bias tuned by  magnetic flux in the main
loop. Energy-level splitting $E_q=\sqrt{\Delta^2+\epsilon^2}$, and
it will be useful for the following analysis to define an angle
$\theta$ by $\tan\theta={\epsilon}/{\Delta}$. $\hat{p}$ and
$\hat{x}$ are the momentum operator and position operator of the
harmonic oscillator. $\hat{\sigma_{x_i}}$ and $\hat{\sigma_{z_i}}$
are the Pauli operators of the $i$th qubit.

We express the Hamiltonian of the oscillator and interaction using
$\hat{a}^\dag$ and $\hat{a}$ (creation and annihilation operator of
the oscillator), so $\hat{H}_{os}$ and $\hat{H}_{int}$ are rewritten
as
\begin{equation}\label{another expression}
 \begin{aligned}
  \hat{H}_{os}=\hbar\omega_0\hat{a}^\dag\hat{a}+\frac{1}{2}\hbar\omega_0,  \\
  \hat{H}_{int}=\lambda(\hat{a}^\dag+\hat{a})(\hat{\sigma}_{z_1}+\hat{\sigma}_{z_2}),  \\
  \lambda=\sqrt{\frac{\hbar}{2m\omega_0}}g.
\end{aligned}
\end{equation}

There is no analytic solution to Eq. (\ref{total Hamiltonian}) so
far, but some approximation methods have been discussed for
one-qubit case without RWA (such
as~\cite{Irish-2007-PRL},~\cite{Casanova-2010-PRL}). In the next
section, we present two adiabatic approximations that can be
used to describe our system under different regimes of parameters.

\section{adiabatic approximations under two opposite extreme conditions}

\subsection{Adiabatic approximation in the displaced oscillator basis}

In this section we consider the case in which $\hbar\omega_0$ is far
greater than $E_q$. In this case, using adiabatic approximation, one
can consider that each of the qubits has a well-defined value of
$\sigma_z$, i.e., $\sigma_{z_1}=\pm1$, and
$\sigma_{z_2}=\pm1$~\cite{Irish-2005-PRB}. When
$\sigma_{z_1}=\sigma_{z_2}=\pm1$, eigenstates of the system can be
written as
$|\psi_\pm,\pm\rangle=|\psi_\pm\rangle\otimes|\pm\rangle$,
where$|\psi_\pm\rangle$ stand for eigenstates of the oscillator,
$|+\rangle$ and $|-\rangle$ stand for the qubits' eigenstates
$|e_1,e_2\rangle$ and $|g_1,g_2\rangle$, respectively. When
$\sigma_{z_1}=-\sigma_{z_2}=\pm1$, eigenstates of the system can be
written as $|\psi_0,0\rangle=|\psi_0\rangle\otimes|0\rangle$, where
$|0\rangle$ stands for $|e_1,g_2\rangle$ and $|g_1,e_2\rangle$.

In the case $\sigma_{z_1}=\sigma_{z_2}=\pm1$ which means the states
of qubits are $|e_1,e_2\rangle$ or $|g_1,g_2\rangle$, the effective
Hamiltonian of the oscillator ~\cite{Ashhab-2010-PRA}
\begin{equation}\label{1 oscillator effective Hamiltonian 1}
\hat{H}_{os,eff}|_{\sigma_{z_1}=\sigma_{z_2}=\pm1}=\hbar\omega_0\hat{a}^\dag\hat{a}\pm2\lambda(\hat{a}+\hat{a}^\dag).
\end{equation}

Assuming $\lambda$ and $\omega_0$ are all real, eigenstates of this
Hamiltonian are as follows which can be viewed as displaced Fock
states:
\begin{equation}\label{1 eigen states 1}
|\psi_\pm\rangle=e^{\mp(2\lambda/\hbar\omega_0)(\hat{a}^\dag-\hat{a})}|n\rangle=|n_\pm\rangle,
\ \ \ \ \ \ \ \ \ \ n=0,1,2,\ldots,
\end{equation}
The eigenenergies are given by
\begin{equation}\label{1 eigen energies 1}
E_{n_\pm}^\prime=n\hbar\omega_0-4\lambda^2/\hbar\omega_0,\ \ \ \ \ \
\ \ \ \ n=0,1,2,\ldots.
\end{equation}

In the case $\sigma_{z_1}=-\sigma_{z_2}=\pm1$ which means states of
qubits are $|e_1,g_2\rangle$ or $|g_1,e_2\rangle$, the effective
Hamiltonian of the oscillator is the same as the usual harmonic
oscillator
\begin{equation}\label{1 oscillator effective Hamiltonian 2}
\hat{H}_{os,eff}|_{\sigma_{z_1}=-\sigma_{z_2}=\pm1}=\hbar\omega_0(\hat{a}^\dag\hat{a}+1/2).
\end{equation}
So the eigenstates are given by Fock states
\begin{equation}\label{1 eigen states 2}
|\psi_0\rangle=|n\rangle=|n_0\rangle, \ \ \ \ \ \ \ \ \ \
n=0,1,2,\ldots,
\end{equation}
and eigenenergies are given by
\begin{equation}\label{1 eigen energies 2}
E_{n_0}^\prime=\hbar\omega_0(n+1/2)\ \ \ \ \ \ \ \ \ \
n=0,1,2,\ldots.
\end{equation}

\begin{figure}[htb]
\centerline{\includegraphics[width=10cm]{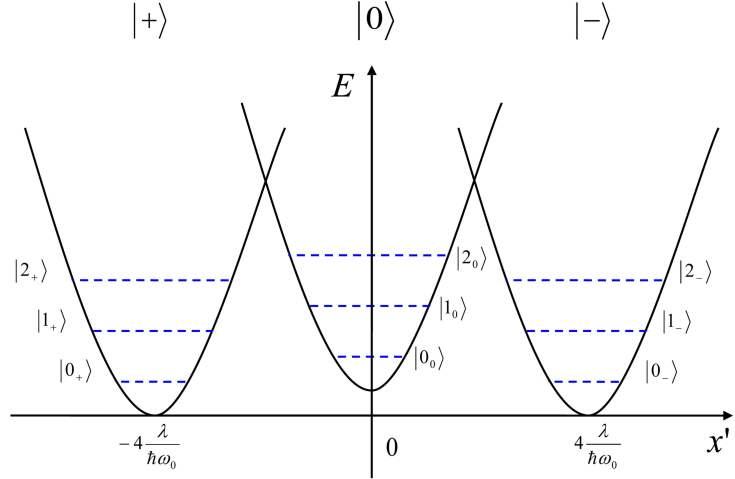}} \caption{(Color
online) Schematic diagram of the displaced oscillator basis. The
horizontal axis $x^\prime=x\sqrt{\frac{2m\omega_0}{\hbar}}$. All
three wells maintain the same harmonic character, and usual
eigenstates as well. The equilibrium position of the left (or the
right) well is shifted by a specific constant. The shift direction
is to the left (or right) when the qubits are in
$|+\rangle=|e_1,e_2\rangle$ (or $|-\rangle=|g_1,g_2\rangle$). The
middle potential well which is double degenerate corresponds to
non-displaced case in which the states of the two qubits are
opposite, i.e., $|0\rangle$, and the equilibrium position is higher
than the others. Eigenstates which have the same value of $n$ in the
left are degenerate with the right well in energy.}\label{Fig2}
\end{figure}

The eigenstates and eigenenergies obtained by Eqs. (\ref{1 eigen
states 1}), (\ref{1 eigen energies 1}), (\ref{1 eigen states 2}),
and (\ref{1 eigen energies 2}) build the main results of displaced
oscillator basis, which will be used throughout the following
analysis. The potentials corresponding to above results are harmonic
oscillator potentials which are illustrated by Fig. \ref{Fig2}.

Because the eigenstates of the oscillator depend on the states of
qubits now, there will be some new properties of the oscillator's
states. States of the same potential well (presented in Fig.
\ref{Fig2}) still maintain the usual orthonormality, that is
$\langle m_+|n_+\rangle=\delta_{mn}$, $\langle
m_0|n_0\rangle=\delta_{mn}$, $\langle m_-|n_-\rangle=\delta_{mn}$,
while states of different potential wells do not, the displacement
operator cause a displacement in $x$, so they are no longer
orthogonal to each other. The overlaps between displaced oscillator
basis of different wells are given by
\begin{equation}\label{1 overlap 1}
\langle m_-|n_0\rangle=
 \left\{\begin{aligned}
  e^{-2\lambda^2/\hbar^2\omega_0^2}\left(-2\lambda/\hbar\omega_0\right)^{m-n}\sqrt{n!/m!}
    L_n^{m-n}\left[\left(2\lambda/\hbar\omega_0\right)^2\right],  \ \ \ \ m\geq n, \\
  e^{-2\lambda^2/\hbar^2\omega_0^2}\left(2\lambda/\hbar\omega_0\right)^{n-m}\sqrt{m!/n!}
    L_m^{n-m}\left[\left(2\lambda/\hbar\omega_0\right)^2\right],  \ \ \ \ m< n, \\
 \end{aligned}\right.\
\end{equation}

\begin{equation}\label{1 overlap 2}
\langle m_0|n_+\rangle=
 \left\{\begin{aligned}
  e^{-2\lambda^2/\hbar^2\omega_0^2}\left(-2\lambda/\hbar\omega_0\right)^{m-n}\sqrt{n!/m!}
    L_n^{m-n}\left[\left(2\lambda/\hbar\omega_0\right)^2\right],  \ \ \ \ m\geq n, \\
  e^{-2\lambda^2/\hbar^2\omega_0^2}\left(2\lambda/\hbar\omega_0\right)^{n-m}\sqrt{m!/n!}
    L_m^{n-m}\left[\left(2\lambda/\hbar\omega_0\right)^2\right],  \ \ \ \ m< n, \\
 \end{aligned}\right.\
\end{equation}
and
\begin{equation}\label{1 overlap 3}
\langle m_-|n_+\rangle=
 \left\{\begin{aligned}
  e^{-4\lambda^2/\hbar^2\omega_0^2}\left(-4\lambda/\hbar\omega_0\right)^{m-n}\sqrt{n!/m!}
    L_n^{m-n}\left[\left(4\lambda/\hbar\omega_0\right)^2\right],  \ \ \ \ m\geq n, \\
  e^{-4\lambda^2/\hbar^2\omega_0^2}\left(4\lambda/\hbar\omega_0\right)^{n-m}\sqrt{m!/n!}
    L_m^{n-m}\left[\left(4\lambda/\hbar\omega_0\right)^2\right],  \ \ \ \ m< n, \\
 \end{aligned}\right.\
\end{equation}
where $L_m^n$ are the associated Laguerre polynomials. It is worth
noting that $\langle m_0|n_-\rangle=(-1)^{m-n}\langle
m_-|n_0\rangle$, and $\langle m_+|n_0\rangle=(-1)^{m-n}\langle
m_0|n_+\rangle$, which are useful identities in the later
calculation.

Having obtained the eigenstates of the displaced oscillator together
with their properties, we now focus on the qubits. Taking any
specific value of $n$, one can build an effective Hamiltonian of the
qubits for that value of $n$. Due to the fact that for each value of
$n$ there are four qubits' states, namely $|e_1,e_2\rangle$ ,
$|e_1,g_2\rangle$ , $|g_1,e_2\rangle$ , and $|g_1,g_2\rangle$, the
matrix of the effective Hamiltonian of qubits is a $4 \times 4$ matrix in
the space defined by $|n_+,e_1,e_2\rangle$ , $|n_0,e_1,g_2\rangle$ ,
$|n_0,g_1,e_2\rangle$ , and $|n_-,g_1,g_2\rangle$. Based on the
overlaps between displaced oscillator basis obtained previously, we
can calculate the elements of this matrix immediately, and the
matrix is given by
\begin{equation}\label{1 qubit effective Hamiltonian}
\mathbf{\hat{H}_{q,eff}}=\left(
  \begin{array}{cccc}
  -\epsilon &  -\frac{\Delta}{2}\langle n_+|n_0\rangle & -\frac{\Delta}{2}\langle n_+|n_0\rangle & 0 \\
    -\frac{\Delta}{2}\langle n_0|n_+\rangle & 0 & 0 & -\frac{\Delta}{2}\langle n_0|n_-\rangle\\
      -\frac{\Delta}{2}\langle n_0|n_+\rangle & 0 & 0 & -\frac{\Delta}{2}\langle n_0|n_-\rangle\\
  0 & -\frac{\Delta}{2}\langle n_-|n_0\rangle & -\frac{\Delta}{2}\langle n_-|n_0\rangle & \epsilon\\
  \end{array}
\right),
\end{equation}
where $\langle n_+|n_0\rangle$, $\langle n_0|n_+\rangle$, $\langle
n_-|n_0\rangle$ and $\langle n_0|n_-\rangle$ are calculated to be of
the same value as
$e^{-2(\lambda/\hbar\omega_0)^2}L_n\left(4\lambda^2/\hbar\omega_0^2\right)$.

Eigenenergies of this effective Hamiltonian are given by
\begin{equation}\label{1 eigen energies qubit}
 \begin{aligned}
    E_{n_\pm}^{\prime\prime}&=\pm\sqrt{\epsilon^2+\Delta^2\left[e^{-2(\lambda/\hbar\omega_0)^2}L_n\left(4\lambda^2/\hbar\omega_0^2\right)\right]^2},\\
    E_{n_0}^{\prime\prime}&=0,\\
    n&=0,1,2,\ldots.\\
 \end{aligned}
\end{equation}
and eigenstates are given by
\begin{equation}\label{1 eigne states system}
 \begin{aligned}
    |\psi_{n_-}\rangle&=\left(1+2\Theta_n^2+2\Theta_n\sqrt{1+\Theta_n^2}\right)|n_+,e_1,e_2\rangle+|n_-,g_1,g_2\rangle \\
                      &\ \ \ +\left(\Theta_n+\sqrt{1+\Theta_n^2}\right)|n_0,e_1,g_2\rangle+\left(\Theta_n+\sqrt{1+\Theta_n^2}\right)|n_0,g_1,e_2\rangle, \\
    |\psi_{n_{0_1}}\rangle&=-|n_+,e_1,e_2\rangle+|n_-,g_1,g_2\rangle+2\Theta_n|n_0,e_1,g_2\rangle,\\
    |\psi_{n_{0_2}}\rangle&=-|n_0,e_1,g_2\rangle+|n_0,g_1,e_2\rangle,\\
    |\psi_{n_+}\rangle&=\left(1+2\Theta_n^2-2\Theta_n\sqrt{1+\Theta_n^2}\right)|n_+,e_1,e_2\rangle+|n_-,g_1,g_2\rangle \\
                      &\ \ \ +\left(\Theta_n-\sqrt{1+\Theta_n^2}\right)|n_0,e_1,g_2\rangle+\left(\Theta_n-\sqrt{1+\Theta_n^2}\right)|n_0,g_1,e_2\rangle, \\
 \end{aligned}
\end{equation}
where
$\Theta_n=\frac{\tan\theta}{e^{-2(\lambda/\hbar\omega_0)^2}L_n\left(4\lambda^2/\hbar\omega_0^2\right)}$.

\begin{figure}[htb]
\centerline{\includegraphics[width=12cm]{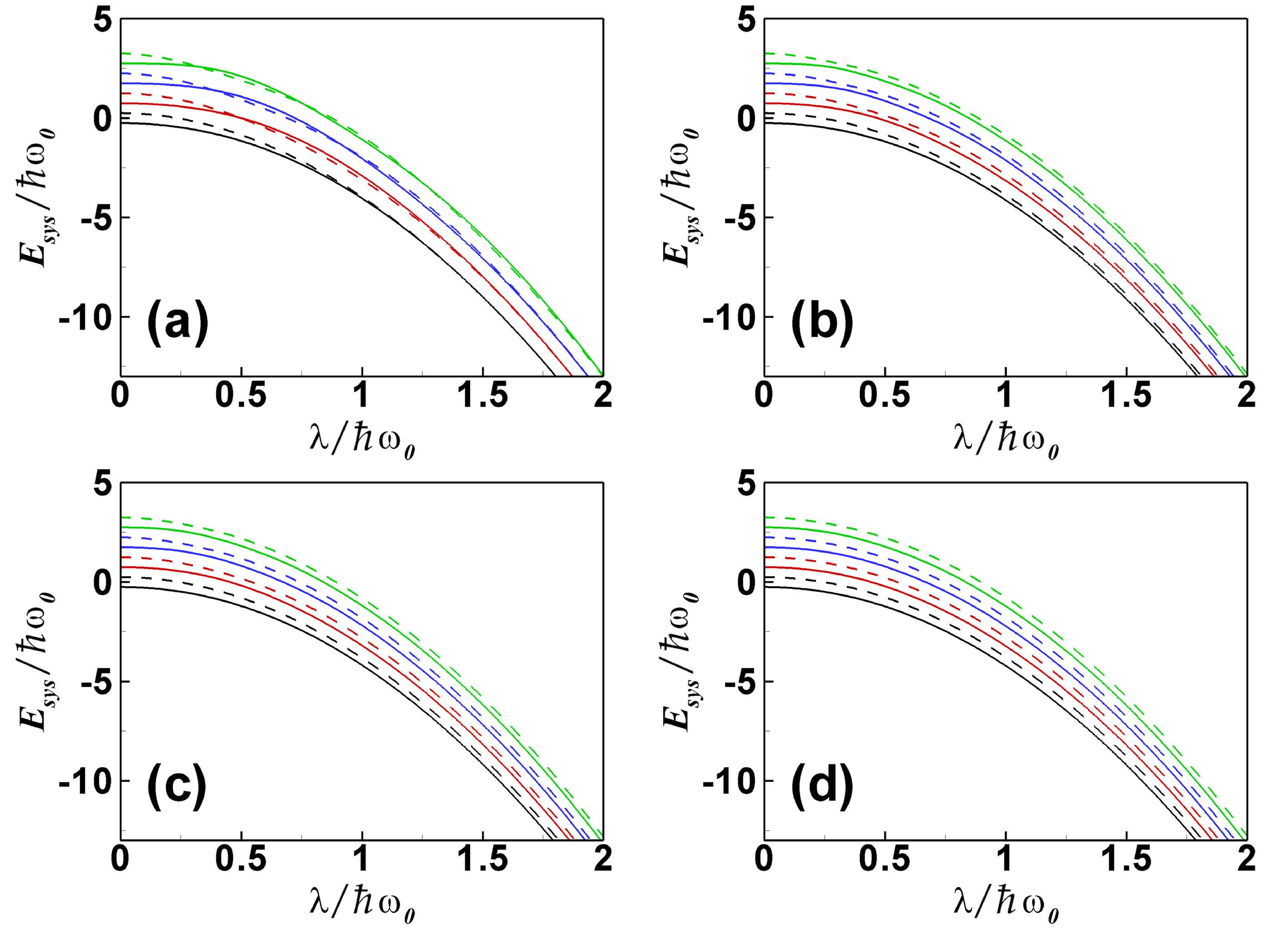}} \caption{(Color
online) The energy-level spectrum diagram of several lower levels as
a function of $(\lambda/\hbar\omega_0)$. $\hbar\omega_0/E_q=4$. The
different color denotes different states of the oscillator(black for
$n=0$, red for $n=1$ , blue for $n=2$ , and green for $n=3$). The
solid lines stand for qubits in state $|g_1,g_2\rangle$, and the
dashed lines stand for $|e_1,e_2\rangle$. (a) $\theta=0$, in the
limit of large $\lambda$ the levels which have the same value of $n$
form pairs. (b) $\theta=\pi/6$, avoided crossing vanishes and
splitting of energy-level pairs with the same value of $n$ turns up
in the large $\lambda$ limit. (c) $\theta=\pi/4$, splitting of
energy pairs enhances. (d) $\theta=\pi/3$, there is no obvious
energy pairs any more, and space between each energy levels
maintains the same in the duration of increasing $\lambda$.
}\label{Fig3}
\end{figure}

The energies of the system are drawn as
$E_{n_\pm}=E_{n_\pm}^\prime+E_{n_\pm}^{\prime\prime}$ and
$E_{n_0}=E_{n_0}^\prime+E_{n_0}^{\prime\prime}$. We present the
energy-level spectrum  as a function of
$\lambda/\hbar\omega_0$ in Fig. \ref{Fig3}, and these four pictures
are different in $\theta$. It is obvious that no matter how we set
the regime of the parameters, $E_{n_0}$ remain constant for any
specific value of $n$, so we do not present these horizontal lines
in the figures. All of the four diagrams are under the condition
that $\hbar\omega_0/E_q=4$ to ensure the adiabatical approximation.
When $\lambda=0$, energy levels of the system are simplified to
$n\hbar\omega+E_q$ which are different from the energy-levels of the
usual harmonic oscillator by a constant, so they are equally spaced.
As $\lambda$ increases, behaviors are evidently different for
different value of $\theta$. In the case $\theta=0$ (i.e. qubits are
in the degenerate point), when $\lambda$ increases, some slight
avoided crossing emerges, and in the limit of large $\lambda$ the
levels which have the same value of $n$ form pairs. In the case
$\theta=\pi/6$, avoided crossing vanishes and splitting of
energy-level pairs with the same value of $n$ turns up in the large
$\lambda$ limit. The splitting of energy pairs enhances when
$\theta=\pi/4$. In the case $\theta=\pi/3$, there is no obvious
energy pairs any more, and space between each energy levels maintain
the same in the duration of increasing $\lambda$.

\begin{figure}[htb]
\centerline{\includegraphics[width=10cm]{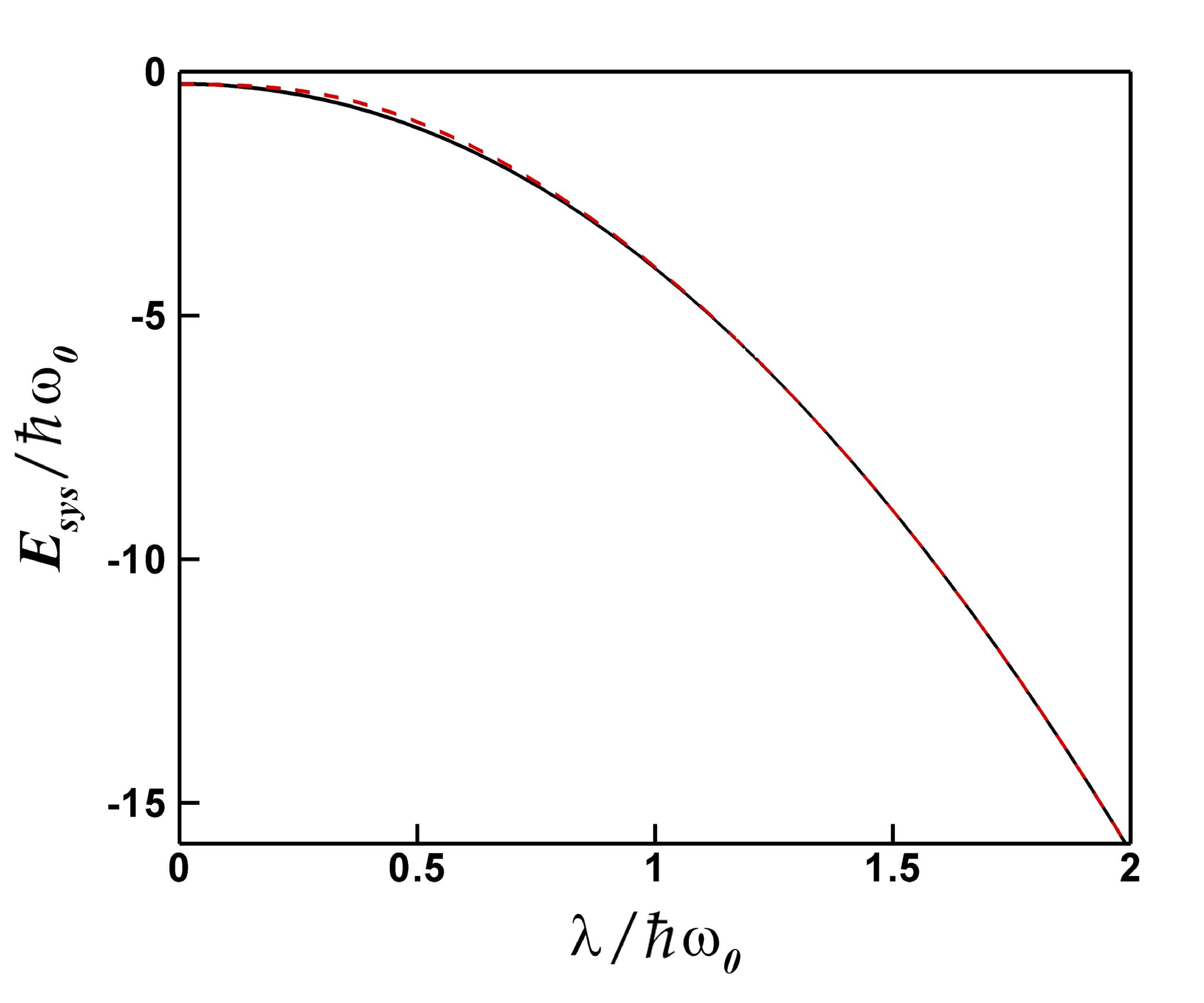}} \caption{(Color
online) The comparison between the displaced oscillator adiabatical
approximation and the numerical solution for the lowest energy
level. They fit well in almost all regime of $\lambda$.
}\label{Fig4}
\end{figure}

We show the comparison between the displaced oscillator adiabatical
approximation method which is given by Eq. (\ref{1 eigen energies
qubit}) and the numerical solution for several lower levels in
Fig. \ref{Fig4}. They fit well in almost all regime of
$\lambda$.

When the qubits are in the degeneracy point, i.e., $\epsilon=0$, the
states of the system coresponding to the eneigys $E_{n_-}$,
$E_{n_0}$ (double degenerate) and $E_{n_+}$ are superposition states
of the displaced basis, which are given by
\begin{equation}\label{1 special eigen states}
 \begin{aligned}
    |\psi_{n_-}\rangle&=1/4|n_+,e_1,e_2\rangle+1/4|n_0,e_1,g_2\rangle+1/4|n_0,g_1,e_2\rangle+1/4|n_-,g_1,g_2\rangle,\\
    |\psi_{n_{0_1}}\rangle&=-1/\sqrt{2}|n_+,e_1,e_2\rangle+1/\sqrt{2}|n_-,g_1,g_2\rangle,\\
    |\psi_{n_{0_2}}\rangle&=-1/\sqrt{2}|n_0,e_1,g_2\rangle+1/\sqrt{2}|n_0,g_1,e_2\rangle,\\
    |\psi_{n_+}\rangle&=1/4|n_+,e_1,e_2\rangle-1/4|n_0,e_1,g_2\rangle-1/4|n_0,g_1,e_2\rangle+1/4|n_-,g_1,g_2\rangle.\\
 \end{aligned}
\end{equation}

It is interesting that no matter how we set the regime of the
parameters, $|\psi_{n_{0_2}}\rangle$ always has the form mentioned
in Eqs. (\ref{1 eigne states system})-(\ref{1 special eigen
states}). It can be rewritten as
$|\psi_{n_{0_2}}\rangle=1/\sqrt{2}(-|e_1,g_2\rangle+|g_1,e_2\rangle)\otimes|n_0\rangle$,
which means that oscillator is decoupled to the states of the
qubits. So the two-qubit maximally entangled states can be obtained by
detecting the state of the oscillator.

\subsection{Adiabatic approximation for the case of high-frequency qubits}

In this section we consider the case that $E_q$ is far larger than
$\hbar\omega_0$. In this case, similar to what we have discussed
above, using adiabatic approximation one can consider that the
oscillator has a well-defined value of $x$~\cite{Irish-2005-PRB}.
Thus the effective Hamiltonian of the qubits is ~\cite{Ashhab-2010-PRA}

\begin{equation}\label{2 qubit effective Hamiltonian}
\hat{H}_{q,eff}|_x=-\frac{\Delta}{2}(\hat{\sigma}_{x_1}+\hat{\sigma}_{x_2})-\frac{\epsilon}{2}(\hat{\sigma}_{z_1}
                    +\hat{\sigma}_{z_2})+gx(\hat{\sigma}_{z_1}+\hat{\sigma}_{z_2}).
\end{equation}
The eigenenergies of Eq. (\ref{2 qubit effective Hamiltonian}) are
given by
\begin{equation}\label{2 eigen energies}
 \begin{aligned}
    E_{q_\pm}&=\pm\sqrt{\Delta^2+(2gx-\epsilon)^2},\\
    E_{q_0}&=0, \ \ \ \text{(double\ degenerate)}\\
 \end{aligned}
\end{equation}

Unlike above results that energies of high-frequency
oscillator are independent of the qubits' states, the energies of
high-frequency qubits now have a dependence on the position $x$ of
the oscillator. Thus the effective potential of the oscillator has
the form
\begin{equation}\label{2 effective potential}
 \begin{aligned}
    V_{os_\pm}&=\frac{1}{2}m\omega_0^2x^2\pm\sqrt{\Delta^2+(2gx-\epsilon)^2},\\
    V_{os_0}&=\frac{1}{2}m\omega_0^2x^2,\\
 \end{aligned}
\end{equation}
where $V_{os_-}$, $V_{os_+}$ and $V_{os_0}$ corespond to the qubits
in states $|g_1,g_2\rangle$, $|e_1,e_2\rangle$, and
$|e_1,g_2\rangle$ (or $|g_1,e_2\rangle$), respectively.

Due to the correction terms in $V_{os_\pm}$, when the qubits are in
the same state, the effective potential of the oscillator is no
longer harmonic, while it is the usual harmonic when the qubits are
in opposite states. However , we can sitll obtain an approximate
solution similarly to Ref.~\cite{Ashhab-2010-PRA}

A renormalized frequency $\tilde{\omega}_0$ is obtained, which is
given by
\begin{equation}\label{2 renormalized frequency}
 \begin{aligned}
    \tilde{\omega}_{0_\pm}^2&=\omega_0^2\pm4g^2/mE_q,\\
    \tilde{\omega}_{0_0}^2&=\omega_0^2,\\
 \end{aligned}
\end{equation}
thus the relevant approximate effective potential is given by
\begin{equation}\label{2 approximate effective potential}
 \begin{aligned}
    V_{os_\pm}&\approx\frac{1}{2}m\tilde{\omega}_{0_\pm}^2\left(x\mp\frac{2\epsilon g}{m\tilde{\omega}_{0_\pm}^2E_q}\right)^2\pm E_q,\\
    V_{os_0}&=\frac{1}{2}m\tilde{\omega}_{0_0}^2x^2,\\
 \end{aligned}
\end{equation}
It is interesting that the oscillator's frequency now has a
dependence on the qubits' states, it increases when both of the
qubits are in ground states, decreases when both of the qubits are
in excited states, and stays unchanged when the qubits are in
opposite states.

For $V_{os_+}$, i.e., the qubits are both in excited states, the
approximate effective potential is a displaced harmonic potential
with an increased frequency, so the states of the oscillator are
displaced Fock states discussed in Sec. III.A. For $V_{os_0}$, i.e.,
the qubits are in opposite states, the effective potential is a
usual harmonic potential, thus the oscillator's states are Fock
states.

It is worthy to note that for $V_{os_-}$, the renormalized frequency
$\tilde{\omega}_{0_-}$ turns into imaginary when
$\omega_0^2<4g^2/mE_q$ (the stationary point), which means that
under the condition
\begin{equation}\label{2 critical point}
\frac{m\omega_0^2E_q}{4g^2}<1
\end{equation}
the system becomes unstable, and the approximate effective potential
is not applicative any more as the effective potential becomes
double-well. In the case of degeneracy point ($\epsilon=0$), by
differentiating $V_{os_-}$ given in Eq. (\ref{2 effective
potential}), one can obtain that the two minimal points of the
double-well potential are located at $\pm x_0$, where
$x_0=\sqrt{4g^2/m^2\omega_0^4-\Delta^2/4g^2}$. The energy barrier
height between the two well is determined by $g^2/m\omega_0^2$. The
effect of introducing a finite value of $\epsilon$ is that two wells
are no longer symmetrical and the location of energy barrier is
shifted to the left or right.

\section{conclusion}

In this paper we present a system that consists of two flux qubits
coupled strongly to a transmission line oscillator, and obtain
eigenenergies of the system and the properties of its eigenstates.
Adiabatic approximation methods under two opposite extreme
conditions, i.e., the adiabatic approximation in the displaced
oscillator basis and the adiabatic approximation of the
high-frequency qubits, which can be used to analyse our system are
compared. Although they start by different assumptions, it has been
proved that they are both valid in the ultrastrong coupling regime.
There are some differences between these two approximations. It is
notable that unlike the later approximation, there is no stationary
point that turns harmonic potential into double-well potential in
the former.

\bigskip

\noindent \textit{Acknowledgments -}  This work was supported by the NSFC (Grant No.
60978009), the Major Research Plan of the NSFC (Grant No. 91121023 ), and the SKPBR of China
(Grant No.2011CBA00200 ).

\end{document}